\documentclass[fleqn,11pt]{article}
\usepackage{amsmath}
\usepackage{amssymb}

\usepackage{amsfonts,amssymb,cite}
\usepackage{graphicx}

\def\noi{\noindent}
\def\jnumber#1#2{\thispagestyle{empty} \noi\unitlength=1mm
    	\begin{picture}(178,10)
                   \end{picture}}

\newcommand{\Title}[1]{\noi {{\Large\bf #1}}\\[1ex]}

\newcommand{\Author}[2]{\noi{\bf #1}\\[2ex]\noi{\normalsize\it #2}\\}

\newcommand{\Abstract}[1]{\vskip 2mm \begin{center}
        \parbox{16.4cm}{\small\noi #1} \end{center}\medskip}

\newcommand{\foom}[1]{\protect\footnotemark[#1]}

\def\nqq{\hspace*{-2em}}
\def\nhq{\hspace*{-0.5em}}

\def\cm{\hspace*{1cm}}

\def\Jl#1#2{#1 {\bf #2},\ }

\def\ApJ#1 {\Jl{Astroph. J.}{#1}}
\def\CQG#1 {\Jl{Class. Quantum Grav.}{#1}}
\def\DAN#1 {\Jl{Dokl. AN SSSR}{#1}}
\def\GC#1 {\Jl{Grav. Cosmol.}{#1}}
\def\GRG#1 {\Jl{Gen. Rel. Grav.}{#1}}
\def\JETF#1 {\Jl{Zh. Eksp. Teor. Fiz.}{#1}}
\def\JETP#1 {\Jl{Sov. Phys. JETP}{#1}}
\def\JHEP#1 {\Jl{JHEP}{#1}}
\def\JMP#1 {\Jl{J. Math. Phys.}{#1}}
\def\NPB#1 {\Jl{Nucl. Phys. B}{#1}}
\def\NP#1 {\Jl{Nucl. Phys.}{#1}}
\def\PLA#1 {\Jl{Phys. Lett. A}{#1}}
\def\PLB#1 {\Jl{Phys. Lett. B}{#1}}
\def\PRD#1 {\Jl{Phys. Rev. D}{#1}}
\def\PRL#1 {\Jl{Phys. Rev. Lett.}{#1}}

\def\lal{&&\nqq {}}
\def\eq{Eq.\,}
\def\eqs{Eqs.\,}
\def\beq{\begin{equation}}
\def\eeq{\end{equation}}
\def\bear{\begin{eqnarray}}
\def\bearr{\begin{eqnarray} \lal}
\def\ear{\end{eqnarray}}
\def\earn{\nonumber \end{eqnarray}}

\def\nnn{\nonumber\\ \lal }

\def\yy{\\[5pt] {}}
\def\yyy{\\[5pt] \lal }

\def\d{\partial}

\def\const{{\rm const}}

\def\ep{\epsilon}

\begin{document}
\twocolumn[
\jnumber{3}{2017}

\Title{A thin dust shell falling to a Reissner-Nordstr\"om black hole\yy as seen by a freely falling observer}

\Author{Alexander Shatskiy\foom 1}{shatskiyalex@gmail.com}


\Abstract
  {We study the observation of a thin dust shell, radially freely falling to a Reissner-Nordstr\"om
    black hole, by an observer who is also freely and radially falling into this black hole.
    Considered and resolved are several common paradoxes and fallacies peculiar for such problems.
  The results of this analytical study are written as a numerical code that allows for calculating all
  related effects of this model. The numerical result have been presented in a few synthesized videos,
  making a colorful, quantitative and detailed description of the occurring astrophysical phenomena,
  both above and below the horizon.
}

] 

\section{Introduction}
\label{sh-introduction}

  Recently, my colleagues repeatedly encountered with misunderstanding and confusion in matters
  of hypothetically-possible observations of some effects from black holes. Moreover, this
  misunderstanding was sometimes originated by even professional astrophysicists.
  This concerns visual effects which accompany an observer freely falling into a black hole.

  Of course, all these arguments are pure fantasy, in the sense of a possible technical implementation
  of such observations, but no doubt this subject is of methodological interest.
  This interest is related to understanding important effects of general relativity (GR) which accompany
  such hypothetic observations. This led to the idea to write an article in which these hypothetic
  observations would be described in detail and, most importantly, correctly.

  In addition to the methodological study, description and calculation of this model, I have also made some
  virtual animations in which I tried to show exactly the real visual effects that must be seen by a freely
  falling observer (in theory). These videos are under open internet access. So, this work (with the videos)
  can be used for effective student learning as well as for all those interested in the subject.

  Briefly about the model:

  1. Let us imagine an observer freely falling into a black hole in vacuum, along the radius,
  and he is always in weightlessness.

   2. In addition, a dust sphere fall into the black hole too, and the observer sees this sphere,
   which is closer to the black hole than the observer.

  3. All dust particles of this sphere are at the same distance from the black hole at the same time.
   All dust particles also fall radially and freely.

  4. The influence of the dust particles and the observer on the black hole is neglected.

  A few words on the misconceptions.
  The main misconception, to be refuted in this paper, is that the observer should see an infinite red shift
  from the dust particles when the sphere reaches the black hole horizon.

  The next misconception is that the observer will no longer see the radiation (the intensity tends to zero)
  when the sphere reaches the black hole horizon.

  Let me briefly explain the cause of the above misconceptions. For photons near the horizon
  it is, in a sense, difficult to escape to infinity, it requires more time than far away from the horizon.
  Therefore an observer flying closer and closer to the horizon sees the photons that broke off the shell
  when it was only approaching the horizon, but the shell itself is at the same time already below the horizon.
  After crossing the black hole horizon, the observer falls to the center\footnote
          {Here and below, by the term ``center'' I mean a smaller radial coordinate $r$ than
            the location discussed.}
  faster than for the photons from the dust, which are still ``trying to get out,'' but gravity ``pulls''
  them to the center.  And the observer, during the fall,  still ``comes across'' these photons.
  For this reason the observer will not see any peculiarities when observing the photons from the shell,
  everything will look smooth and continuous!

  Another common misconception is that under the horizon of a black hole we cannot use the
  Schwarzschild or Reissner-Nordstr\"om static coordinates --  allegedly  because below the horizon
  space and time interchange. The latter statement is fundamentally wrong: can only say that some
  properties are reversed in time and space, namely, the signs in the metric, but time still remains time
  and space remains space below the horizon.

  Therefore, the static coordinates below the horizon can can be used, otherwise the conclusions on
  the existence of a singularity would be mistaken as well as the conclusions on the existence and
  location of the Cauchy horizon for real black holes. For a correct description of the observational
  effects below the horizon, it is necessary to write all formulas describing these effects in an invariant
  form, as is done in the following sections. As is known, the value of an invariant (or scalar) at a given
  point of space-time dos not depend on the choice of a coordinate system or reference frame.

  In many other studies we have performed such calculations below the horizon in other frames,
  see [1--4]. Therefore, this choice of the coordinate system is not related to simplicity of calculations
  but is rather related to simplicity of interpretation and understanding. If we correctly repeat the further
  calculations, for example, in the comoving reference frame of freely falling matter, the results will be
  the same.

  The Reissner-Nordstr\"om (electrically charged) black hole has been chosen as the original model.
  It is obvious that in space there cannot be objects with a large electric charge, so my choice of
  the Reissner-Nordstr\"om solution is associated with more fundamental reasons than simply a generalization
  of the Schwarzschild solution. It is associated with the fact that the Schwarzschild solution is
  unrealizable in nature: the real black holes always have some rotation, i.e., the real black holes are always
  Kerr black holes.  At the same time, Kerr black holes have a nontrivial topology, fundamentally different
  from that of the Schwarzschild solution.

  Due to the Cauchy horizon in the Kerr geometry, space-time splits into two internal areas, the T-region
  between the horizons and the R2-region under the inner horizon. The dynamics and observational
  manifestations of these interior regions are much more complex, interesting and multifaceted than for the
  single internal T-region in the Schwarzschild solution. However, the examination of the analytical and
  numerical models in the Kerr solution is much more difficult than in the Schwarzschild one.
  In the Kerr metric it is impossible to consider a spherical dust shell: it will be necessary bent and broken
  by gravimagnetic forces which are present in the Kerr solution. The maximum that could be considered
  in the Kerr solution analytically is the dynamics of a thin dust ring which falls and rotates in the equatorial
  plane of the Kerr coordinates. Therefore, the solution of this problem in the Reissner-Nordstr\"om metric
  is a compromise between reality and the complexity of the solution.

  The consideration in the Reissner-Nordstr\"om metric has roughly the same complexity as in the
  Schwarzschild metric. It is the reason for choosing the Reissner-Nordstr\"om black hole, which, as well as
  the Kerr black hole, has a nontrivial topology and an inner Cauchy horizon, so that
  the Reissner-Nordstr\"om solution is closer to reality than the Schwarzschild simplest  solution.

\section{The laws of motion}   \label{sh-s_LOW}

  The law of motion (the relation between time $t$ and the radius $r$) for a spherical thin dust shell,
  freely falls to a Reissner-Nordstr\"om black hole, in the Reissner-Nordstr\"om coordinates can be written
  in the general form
\bear             \label{sh-shell_low0}
            t_{\rm shell}(r) = t_{\rm shell^0} + \int_{r_{\rm shell^0}}^{r} F_{\rm shell}(r)\, dr,
\ear
  where ${F_{\rm shell}(r)}$ is some function to be determined further.

  Similarly, for a photon in the same coordinates in the plane ${(r, \, \theta)}$ we can also write a law
  of motion:
\bear                  \label{sh-photon_low1}
          t_{\rm photon}(r, h) = t_{\rm photon}^{\rm shell}(h) + \int\limits_{r_{\rm shell}}^{r} \nhq
                   F_{\rm photon}(r,h)\, dr.
\ear
  Here $h$ is the impact parameter of the photon which is directly related to the angle
  ${\Theta_{\rm shell}(h)}$ at which the photon was emitted from the shell;
  ${F_{\rm photon}(r,h)}$ is some function for the photon, which will also be determined further.

  Using \eq (\ref{sh-photon_low1}), we obtain the time of radiation for a photon radially emitted from
  the shell with $h=0$ and $\theta_{\rm shell} = 0$:
\bear                \label{sh-photon_low2}
     t_{\rm photon}(r_{\rm shell^0}, 0) = t_{\rm photon}^{\rm shell}(0) = t_{\rm shell^0}
\ear
  and the time for this photon to reach the observer:
\bear              \label{sh-photon_low3}
               t_{\rm photon}(r_{obs^0}, 0) = t_{\rm shell^0} + \!\!\int\limits_{r_{\rm shell^0}}^{r_{obs^0}}\nhq
                    F_{\rm photon}(r,0)\, dr.
\ear
  Since we consider only emitted photons (at different times) which reach the observer at the same time,
  we set this time equal to ${t_{\rm obs^0}}$:
\bearr          \label{sh-photon_low4}
          t_{\rm photon}(r_{\rm obs^0}, h)  = t_{\rm photon}^{\rm shell}(h)
\nnn          \cm
              + \int_{r_{\rm shell}}^{r_{\rm obs^0}} F_{\rm photon}(r,h)\, dr := t_{\rm obs^0}.
\ear
  The time of the radiation (to the observer) of a given photons corresponds to the radius
  $r_{\rm shell}$ of the shell, therefore, according to (\ref{sh-shell_low0}), we also have
\bear                \label{sh-shell_low2}
            t_{\rm shell}(r_{\rm shell}) = t_{\rm shell^0}
                              + \int\_{r_{\rm shell}}^{r_{\rm shell^0}} F_{\rm shell}(r)\, dr.
\ear
  Here we have taken into account that earlier points in time correspond to larger radii and vice versa.
  Therefore, the integral in (\ref{sh-shell_low2}) should be negative because it must satisfy the condition
   ${r_{\rm shell^0}< r_{\rm shell}}$ that a photon with ${h>0}$ has time to reach the observer at the
  same time as a radial photon with ${h=0}$. Since the nonradial photon needs more time it must be
  emitted before (at a larger radius).

  The time ${t_{\rm photon}^{\rm shell}(h)}$ of photon emission from the
  shell coincides (by definition) with the time
   ${t_{\rm shell}(r_{\rm shell})}$, so from (\ref{sh-photon_low4}) and (\ref{sh-shell_low2}) we have
\bearr                \label{sh-photon_shell1}
               t_{\rm photon}^{\rm shell}(h) = t_{\rm obs^0} + \int_{r_{\rm obs^0}}^{r_{\rm shell}}
                       F_{\rm photon}(r,h)\, dr
\nnn
             =   t_{\rm shell^0} - \int_{r_{\rm shell^0}}^{r_{\rm shell}} F_{\rm shell}(r)\, dr.
\ear
  Above, we assumed that the observer is located at radius $r_{\rm obs^0}$ at the moment
  $t_{\rm obs^0}$ of the arrival time of photons. But our observer is freely falling, so his coordinates
  (and the arrival time of photons) are also changing. To account for that, it is sufficient to replace
  all zero indices with the current index (such as the index $k$:
   ${t_{\rm obs^k}, \, r_{\rm obs^k}, \, t_{\rm shell^k}, \, r_{\rm shell^k}}$).
  Then the zero index ${(k=0)}$ will be use as the initial conditions index of our model.

  Let us choose the origin of time readout at ${t_{\rm obs^0} := 0}$.
  Then the integral (\ref{sh-photon_shell1}) can be rewritten as
\bearr                              \label{sh-photon_shell2}
     \int_{r_{\rm shell^k}}^{r_{\rm obs^k}} \left[F_{\rm photon}(r,h) - F_{\rm photon}(r,0)\right]\, dr
\nnn
      = \int_{r_{\rm shell^k}}^{r_{\rm shell}} \left[F_{\rm shell}(r) + F_{\rm photon}(r,h)\right]\, dr.
\ear
  Using this ratio, it is possible to calculate numerically the dependence ${r_{\rm shell}(h)}$ for
  each index $k$ (i.e., for any time point for the observer).

  At the same time, we believe that the time ${t_{\rm shell^k}}$ and the radius ${r_{\rm shell^k}}$
  correspond to the radiation of the photon along the $Z$ axis from the shell during its fall (and the arrival
  time of this photon to the observer at the moment ${t_{\rm obs^k}}$).
  By this time ${t_{\rm obs^k}}$ and by this radius ${r_{\rm obs^k}}$, come all photons which
  have been nonradially emitted by the shell ${(h>0)}$, they are radiated by the shell from radius
  ${r_{\rm shell}}$, at the moments ${t_{\rm shell}}$.

  Differentiating both sides of \eq (\ref{sh-photon_shell2}) in the parameter $h$, we get:
\bearr                      \label{sh-photon_shell3}
            \frac{dr_{\rm shell}}{dh} = \left[ \int_{r_{\rm shell}}^{r_{obk}^k}
                             \frac{hr\, dr}{\left(r^2 - h^2 f\right)^{3/2}}\right]
\nnn \cm \times
                         \left[F_{\rm shell}(r_{\rm shell}) + F_{\rm photon}(r_{\rm shell},h)\right]^{-1}.
\ear
  Hence, we obtain the dependence $r_{\rm shell}(h)$.

\section{Geodesic equations}\label{sh-s_GEO}

  We write the Reissner-Nordstr\"om metric:
\bearr               \label{sh-ds2}
             ds^2 = f(r)\, dt^2 - {dr^2\over f(r)} - r^2(d\theta^2 + \sin^2\theta\, d\varphi^2),
\nnn
              f(r) := \left(1-{r_h\over r}\right)\left(1-{r_c\over r}\right)\, .
\ear
  We use the geodesic equations for a particle moving in this gravitational field (see [5] or [6], \S\ 87):
\bear               \label{sh-GEO}
           \frac{dU_i}{ds}=\frac{1}{2}\frac{\d g_{jl}}{\d x^i} U^j U^l
\ear
  Then for the metric (\ref{sh-ds2}) and for $i$ corresponding to the $t$ coordinate, we have the integral
  of motion: $U_t : = \ep = \const$, and for $i$ corresponding to the $\theta$ coordinate, we have the
  integral of motion ${U_\theta := h \ep = \const}$.

  Given the identity ${U_i U^i\equiv 1}$, we have:
\bearr               \label{sh-u_iu_i}
            U^t = \frac{\ep}{f} \, , \qquad         U^\theta = -\frac{h\ep}{r^2},
\nnn
             \frac{dr}{ds} := U^r = -\sqrt{\ep^2 - f(1+h_\ep^2\ep^2/r^2)} \, ,
\ear
  Here, the minus sign before the root has been chosen according to the direction of motion, towards
  the center. Hence, for the radial fall of a massive particle ${(h_\ep = 0)}$ we get the nonzero
  components ${U^i}$:
\bear          \label{sh-U1}
           U^t = \frac{\ep}{f}\, ,\qquad U^r = -\sqrt{\ep^2 - f} \, .
\ear
  A transition to massless particles is accomplished by replacing in (\ref{sh-u_iu_i})
  ${U^i \to \ep\Psi^i}$ and by the limiting transition ${\ep\to\infty}$:
\bearr               \label{sh-dopler4}
    \Psi^t(r) = \Psi_t /f(r),\qquad            \Psi^\theta (r) = -\frac{h \Psi_t}{r^2},
\nnn
               \Psi^r(r) = \pm \Psi_t\sqrt{1 - h^2 f/r^2}.
\ear
  Here ${\Psi^i}$ is the null 4-vector of the photon: $\Psi^i \, \Psi^j g_{ij} = 0$; the expression
  (\ref{sh-dopler4}) is valid for a single photon, along its entire trajectory, and similarly to massive particles,
  $\Psi_t = \Psi_t(h)$ and $\Psi_\theta(h) := h\Psi_t$ are integrals of motion for the photon.
  Similarly, the expression (\ref{sh-u_iu_i}), plus sign in the expression (\ref{sh-dopler4}) corresponds
  to the direction of photon emission from the center, and the minus sign to the center.

  Knowing the 4-vectors, we obtain the function $F_{\rm shell}(r)$ for massive particles and
  the function $F_{\rm photon}(r,h)$ for photons:
\bearr        \label{sh-photon_shell3}
                  F_{\rm shell}(r) = \frac{1}{f\sqrt{1 - f/\ep^2}} ,
\nnn
     F_{\rm photon}(r,h) = \frac{1}{f\sqrt{1 - h^2 f/r^2}} \ .
\ear

\section{Redshift of a visible shell}       \label{sh-s_RS}

  In the reference frame comoving to a dust particle, the scalar product of the 4-vector ${\psi_i}$ for
  the photon and the velocity 4-vector $u^i := \{1, 0, 0, 0\}$ of the observer is equal to the natural
  frequency of the photon: ${w_{\rm shell} := u^i \psi_i}$. Since the scalar product is an invariant,
  at the same point of the Reissner-Nordstr\"om system, we also have for the radiation frequency:
\bear          \label{sh-dopler3}
               w_{\rm shell} = u^i \psi_i = \Psi_i \, U^i = \Psi_t U^{t} + \Psi_r U^{r}
\ear
  In this case, the components of 4-velocity ${U^i}$ are the components of 4-velocity of dust particles
  in the Reissner-Nordstr\"om coordinates. Taking into account \eq (\ref{sh-dopler4}), the expression
  (\ref{sh-dopler3}) on the dust shell can be rewritten as
\beq  \nhq             \label{sh-dopler5}
    w_{\rm shell} = \Psi_t(h) \! \left[U^{t}_{\rm shell} -
     U^{r}_{\rm shell}\frac{\sqrt{1 - h^2 f_{\rm shell}/r_{\rm shell}^2}}{f_{\rm shell}}\right]
\eeq
  The plus sign in the expression (\ref{sh-dopler4}) for ${\Psi^r}$ has been chosen in accordance
  with the direction of photon emission from the center. Bear in mind that the denominator in the
  expression (\ref{sh-dopler5}) vanishes on the horizon, and the expression becomes singular. This
  corresponds to the fact that the function ${\Psi_t (h)}$ for the emitted photon tends to zero
  (approach of the shell to the horizon). And for a photon directed towards the center
  (the minus sign in (\ref{sh-dopler4})), on the contrary, the singularity is subtracted, and the final
  frequency ${w_{\rm shell}}$ is everywhere finite (as well as the function ${\Psi_t (h)}$).

  Similarly, from \eq (\ref{sh-dopler5}) at the observation point, the frequency measured by the observer is
\bearr                \label{sh-dopler6}
                 w_{\rm obs^k} := \Psi^{\rm obs^k}_i \, U^i_{\rm obs^k} =
                   \Psi_t(h) \Bigg[U^{t}_{\rm obs^k}
\nnn      \cm
                      - U^{r}_{\rm obs^k}\frac{\sqrt{1 - h^2 f_{\rm obs^k}/r_{\rm obs^k}^2}}{f_{\rm obs^k}}
        \Bigg]
\ear
  The Doppler shift $z$ is determined by the frequency ratio ${\rm RS} := w_{\rm shell}/w_{\rm obs^k}$:
\bearr             \label{sh-dopler7}
               1+z := {\rm RS} = \frac{w_{\rm shell}}{w_{\rm obs^k}}
                                = \frac{f_{\rm obs^k}}{f_{\rm shell}}
\nnn  \times
            \frac{1+\sqrt{1-f_{\rm shell}\left[ 1/\ep^2 +h^2(1 - f_{\rm shell}/\ep^2)/r_{\rm shell}^2 \right]}}
                 {1+\sqrt{1-f_{\rm obs^k}\left[ 1/\ep^2 +h^2(1 - f_{\rm obs^k}/\ep^2)/r_{\rm obs^k}^2 \right]}}
\nnn
\ear
  It can be shown that the ratio ${\rm RS} = w_{\rm shell}/w_{\rm obs^k}$ is finite even near the horizon.
  To do that, we note that at approach to the horizon the function $f_{\rm shell}$ becomes small.
  The photon emitted by the shell at a radius $r_{\rm shell}$ is absorbed by the observer at the moment
  $t_{\rm obs^k}$ corresponding to the radius of the observer $r_{\rm obs^k}$.
  And the function $f_{\rm obs^k}$ has the same order of smallness as the function $f_{\rm shell}$
  for the same photon.

  Let us assume (to simplify the calculations) that the value of $\ep$ for the observer is the same as
  for the dust shell, i.e.,  ${F_{\rm obs} := F_{\rm shell}}$. Let us write the following equation:
\bear           \label{sh-ass0}
           t_{\rm obs^k} = t_{\rm shell^k} + \Delta t_{\rm photon^k}
\ear
  Here [see (\ref{sh-photon_low2}), (\ref{sh-photon_low3})]:
\bearr
       t_{\rm obs^k}  = \int_{r_{\rm obs^k}}^{{r_{\rm obs^0}}} F_{\rm shell}\, dr \, , \label{sh-ass1}
\yyy               \label{sh-ass2}
       t_{\rm shell^k}  = t_{\rm shell^0} + \int_{r_{\rm shell^k}}^{{r_{\rm shell^0}}} F_{\rm shell}\, dr \, ,
\yyy
         t_{\rm shell^0}  = -\Delta t_{\rm photon^0}
             = - \nhq \int_{r_{\rm shell^0}}^{r_{\rm obs^0}}\nhq  F_{\rm photon}(r,0)\, dr \, ,\label{sh-ass3}
\yyy      \label{sh-ass4}
    \Delta t_{\rm photon^k} := \int_{r_{\rm shell^k}}^{{r_{\rm obs^k}}} F_{\rm photon}(r,0)\, dr \, .
\ear
  Substituting the integrals (\ref{sh-ass1})--(\ref{sh-ass4}) into \eq (\ref{sh-ass0}) and reducing the intersecting
  areas of integration, we obtain:
\bearr          \label{sh-ass5}
                  T_0 := \int_{r_{\rm shell^0}}^{r_{\rm obs^0}} F_{\rm sum}\, dr
         = \int_{r_{\rm shell^k}}^{r_{\rm obs^k}} F_{\rm sum}\, dr  ,
\nnn
     F_{\rm sum}(r) := F_{\rm shell}(r) + F_{\rm photon}(r,0) \, .
\ear
  The value of $T_0$ determined from the initial conditions has the sense of time in the Reissner-Nordstr\"om
  system. During the time $T_0$ the photon emitted from the shell passes the distance to the observer's
  position at the initial time and then, plus the time in which the observer flies the same distance
  (to the initial shell locations at ${t_{\rm shell^0}}$). However, this does not mean that we will have to
  start numerical integration of our model from the values of the radii ${r_{\rm obs^0}}$ and
  ${r_{\rm shell^0}}$. We might as well begin numerical integration of our model at any values of
  the radii ${r_{\rm obs^k}}$ and ${r_{\rm shell^k}}$ which satisfy \eq (\ref{sh-ass5}).
  Thus the initial assignment of values of the radii ${r_{\rm obs^0}}$ and ${r_{\rm shell^0}}$ determines
  the future relationship between the radii ${r_{\rm obs^k}}$ and ${r_{\rm shell^k}}$ from the integral
  in \eq (\ref{sh-ass5}) for $T_0$.

  If, to both sides of \eq (\ref{sh-ass5}), we add the following difference between the integrals:
\bear
    \int_{r_{\rm obs^0}}^{r_{\rm obs^k}} F_{\rm sum}\, dr
                         - \int_{r_{\rm shell^0}}^{r_{\rm shell^k}} F_{\rm sum}\, dr = 0,    \label{sh-ass5_3}
\ear
  then \eq (\ref{sh-ass5}) doesl not change, so the difference of the integrals in (\ref{sh-ass5_3}) should be zero.

  To determine the redshift value, according to (\ref{sh-dopler7}), it is sufficient to know two
  quantities: ${f_{\rm obs^k}}$ and ${f_{\rm shell^k}}$.
  At the same time, the values of $r_{\rm obs^k} $ and ${f_{\rm obs^k}}$ are set by hand,
  and the values of $r_{\rm shell^k}$ and $f_{\rm shell^k}$ are calculated by the formula (\ref{sh-ass5_3})
  --- we integrate there ${F_{\rm sum}}$ to achieve its zero. The final point of integration in the second
  integral of (\ref{sh-ass5_3}) will be the required radius $r_{\rm shell^k}$ from which we find
  ${f_{\rm shell^k}}$.

  The value of the radius ${r_{\rm shell}}$ required for calculating ${f_{\rm shell}}$ and redshift of a
  photon with an arbitrary impact parameter ${h > 0}$, is calculated by \eq (\ref{sh-photon_shell1}),
  and the quantity ${t_{\rm shell^0}}$ necessary for that is also obtained from \eq (\ref{sh-ass3}) by integration:
\bearr
    \int_{r_{\rm shell}}^{r_{\rm obs^0}} \nhq F_{\rm photon}(r,h)\, dr -
    \int_{r_{\rm shell^0}}^{r_{\rm obs^0}} \nhq F_{\rm photon}(r,0)\, dr
\nnn
          = \int_{r_{\rm shell_0}}^{r_{\rm shell}} F_{\rm shell}(r)\, dr.              \label{sh-get_r_sh}
\ear
  In the limit  ${r_{\rm shell^k} \to r_h}$, ${r_{\rm obs^k} \to r_h}$, ${f_{\rm shell^k} \to 0}$, and
  ${f_{\rm obs^k} \to 0}$, for the function ${F_{\rm sum}(r)}$ we obtain the asymptotic behavior
  $F_{\rm sum}(r) \to 2/f(r)$. Then the integral in (\ref{sh-ass5}) is also has the asymptotic behavior:
\bear
    \int_{r_{\rm shell^k}}^{r_{\rm obs^k}} F_{\rm sum}\, dr \to
    \frac{2r_h^2}{r_h-r_c} \ln\left(\frac{r_{\rm obs^k}-r_h}{r_{\rm shell^k}-r_h}\right).   \label{sh-ass6}
\ear
  Hence we have:
\bear
            \frac{f_{\rm obs^k}}{f_{\rm shell^k}} \to \frac{r_{\rm obs^k}-r_h}{r_{\rm shell^k}-r_h} \to
        \exp\left[\frac{T_0(r_h - r_c)}{2r_h^2}\right].                     \label{sh-ass7}
\ear
  Thus it is clear that the quantity
\[
    f_{\rm obs^k}/f_{\rm shell^k} \to f_{\rm obs^{h+}}/f_{\rm shell^{h+}}
\]
  is finite on the horizon,\footnote
         {Here we denote by the index $h+$ the quantities at which we stop the numerical integration
           before the horizon $r_h$.}
  and therefore the redshift ${\rm RS_h}$ at the horizon is also finite:
\bearr
          {\rm RS_h} = \frac{w_{\rm shell^h}}{w_{\rm obs^h}} \to
        \exp\left[\frac{T_0(r_h - r_c)}{2r_h^2}\right] .                         \label{sh-dopler8}
\ear
  This shows that if $T_0 > 0$ or $r_{\rm obs^0}>r_{\rm shell^0}$, then the frequency shift is
  red at the horizon.

  Conclusion: Due to the choice of the static Reisner-Nordstr\"om coordinate system, we cannot
  pass continuously through the horizons points in our numerical integration.
  But the left and right limits of the expressions (\ref{sh-dopler7}) are the same at the horizon,
  as well as the left and right limits of the ratio ${f_{\rm obs^k}/f_{\rm shell^k}}$,
  therefore we can continue our integration after the horizon with the same limits as before the horizon.

  These and further arguments are perfectly consistent with the membrane paradigm for black holes,
  see [7].

\section{Discussion} \label{sh-s_DISS}

  Since all photons emitted upwards outside the horizon can be also observable only outside the horizons,
  and all photons emitted upwards between the horizons can be also observable only between horizons,
  these two regions are unconnected to each other (in the above sense).

  In the external R1-region (up to falling into the black hole), as shown in (\ref{sh-dopler7}), the redshift
   is determined only by the values of the radii ${r_{\rm shell}}$ and ${r_{\rm obs^k}}$, the point of
  photon radiation and the point of its absorption. You could say that the redshift value is determined by
  the ratio of absolute values of $f(r)$, i.e., on the radii of photon radiation and absorption.
  Therefore, in the external R1-region the radial photon frequency will always decrease (redness).
  As for the nonradial photons (emitted with ${h>0}$), it needs more time (than for the radial photon)
  to reach the observer at the same time at which he/she receives the radial photon.
  Therefore, a nonradial photon must be emitted by the shell earlier, i.e., at a larger radius.
  This can lead to the fact that nonradial photons will experience blue (or violet) frequency shift
  when absorbed by the observer (for sufficiently large values of $h$). But in reality, these ``blue''
  photons correspond to radiation inside by the shell (photons coming down).

\begin{figure*}
\centering
\includegraphics[width=0.85\textwidth]{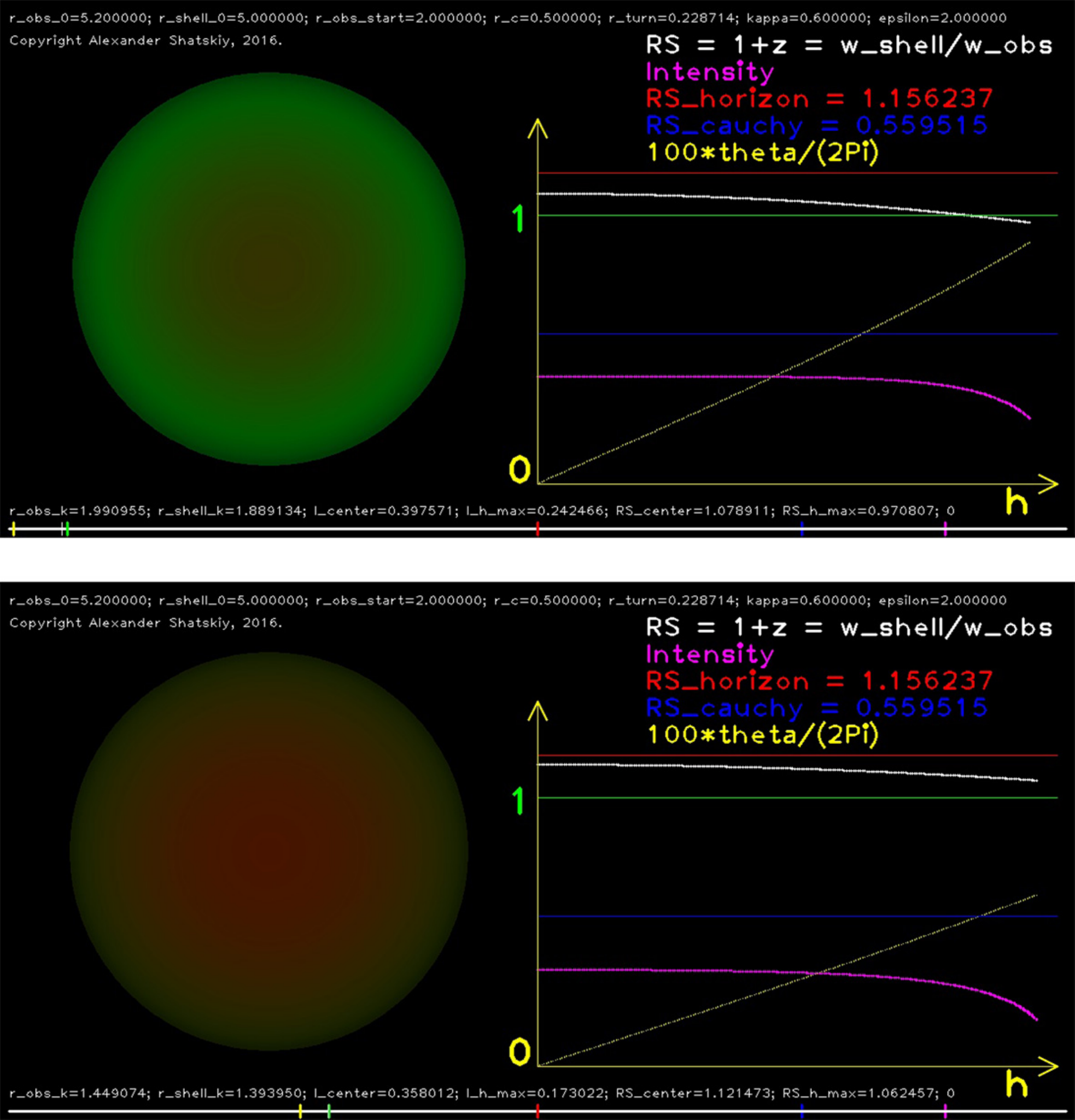}
\caption{\small
    Two video frames for $r_c = 0.5\,r_h$, $\ep = 2$ at $r_{\rm obs^k} = 1.99\,r_h$ and
    $r_{\rm obs^k} = 1.45r_h$, above the horizon.
    On the left there is a black-and-white negative of a colored picture of a falling dust shell for an observer
    falling behind it. Darker points correspond to brighter points in the colored picture. On the right there are
    plots of the photon redshift (RS $= 1+z$), their intensity and their total rotation angle
    ($100\, \theta/(2\pi)$), depending on the photon
    impact parameter $h$. The other figures are organized in a similar manner.
}
\label{sh-R1}
\end{figure*}

  In the T-region (between the horizons), amazing things begin to happen to our model!
  The outward emitted radial photon from the shell is trying to move upwards, but the gravity force is
  so great that it still moves down. After some time, the observer ``catches'' this photon...
  Thus it turns out that in the T-region the observation point has always has a {\it smaller\/}
  radius $r_{\rm obs}$ than that of the emission point ($r_{\rm shell}$). Such ``paradoxes'' are
  only inherent to a T-region between the black hole horizons.
  Still the observer sees a redshift from the radial photons (in the early T-region).

  Since the function ${f(r)}$ is non-monotone in the T-region, the modulus $|f(r)|$ reaches its maximum
  there and then decreases back to zero on the Cauchy horizon $r_c$. Therefore, an instant will come
  when the functions $f(r)$ will be the same for the emission point and the absorption.
  This will correspond to the absence of a frequency shift. After this instant, the situation will change ---
  the observer will begin to see a blueshift of the frequency of radial photons. For nonradial photons the
  situation in this area is less predictable: everything depends on the radius (on the value of $|f(r)|$) from
  which a nonradial photon comes to the observer.

  For the same reasons as in the R1-region, nonradial photons need more time than the radial ones.
  Therefore they should be radiated by the shell at a larger radius (toward the outside horizon $r_h$).
  Therefore (due to a non-monotone function $f(r)$ in the T-region) there can be variants with both
  redshifts and blueshifts of the frequency for nonradial photons.

  However, there is one important remark: at once under the horizon $r_h$, the observer cannot see
  a blueshift for any photons. This is due to the fact that all radiating shells (both for radial and nonradial
  photons) are at large radii (but under $r_h$), therefore the frequency shift will be only red.
  In this regard, immediately above the horizon $r_h$ (in the external R1-region) the frequency shift for
  the observer can only be red too because the observer can see only a continuous change in the frequency
  shift. It is possible that the observer before arriving at the horizon $r_h$ no longer sees such ``blue''
   photons coming to him from inside the shell with sufficiently large values of $h$.

  In the interior R2-region (under the Cauchy horizon $r_c$), the situation once again radically changes.
  Now radial photons radiated up by shell can ``overcome'' the gravity force and move up  to the incident
  observer. The observer will see these radial photons at a larger radius than at the radiation instant,
  but at lower values of the function  ${f(r)}$. It will still conform to the blueshift for the radial photons.
  Now the observer in the R2-region can no longer see the photons emitted upward in the T-region
  because they ``drift'' down only to the Cauchy horizon and then remain on this horizon.
  Therefore, the observer sees only photons from the R2-region,\footnote
         {As was said above, the observer can see the photons (emitted upwards) only from the same region
            in which he is at the moment.}
  and these are photons coming to him from large radii, closer to the Cauchy horizon $r_c$.
  Thus, again, it can happen that a nonradial photon from the R2-region comes to the observer from
  a larger radius than the observer's when watching it. In this case, the observer will see this nonradial
  photon as being red. Thus in the interior R2-region the observer will also be able to see both blue-
  and redshifted photon frequency (for sufficiently large $h$).

\section{Motion between the horizons} \label{sh-s_T}

  All previous arguments and formulas remain the same in the T-region (between the horizons). Therefore,
  in \eqs (\ref{sh-ass1})--(\ref{sh-ass5_3}) it makes sense to replace the indices ``0'' with ``h'', related to the
  horizon $r_h$ (the minus index means ``under the horizon''). In the T-region, the sign of
  the expressions $f$, $F_ {\rm shell}$, and $F_ {\rm photon}$ changes to the opposite, it provides
  the necessary direction for the photons since the photons radiated outwards in this area ``fly'' to the center.

\begin{figure*}
\centering
\includegraphics[width=0.85\textwidth]{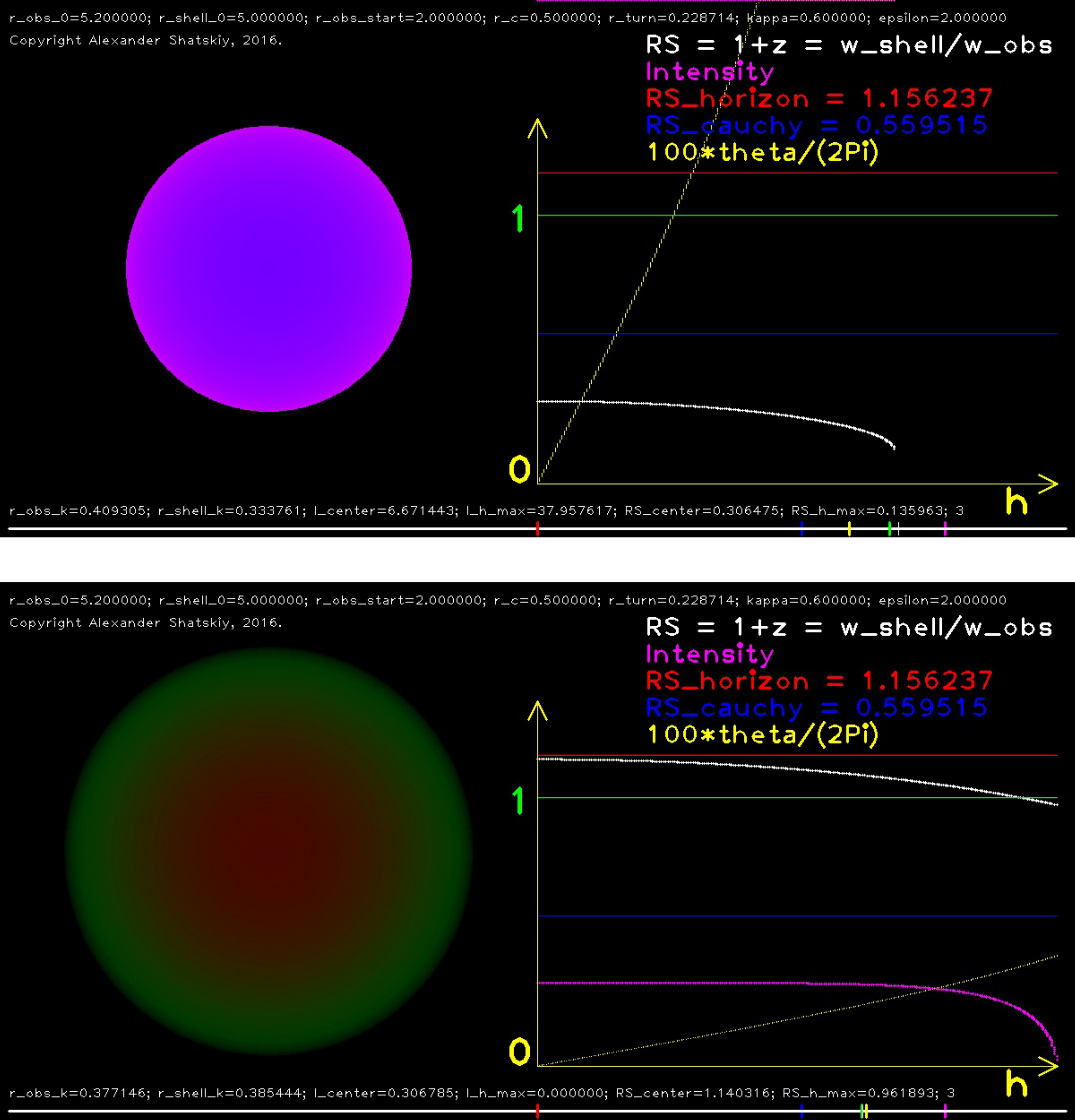}
\caption{\small
           Two video frames for $r_c=0.5\,r_h$, $\ep = 2$ at $r_{\rm obs^k}= 0.88\, r_h$ and
           $r_{\rm obs^k} = 0.66\, r_h$, between the horizons.}       \label{sh-R2}
\end{figure*}
  It is obvious (and verifiable in the the comoving reference frame) that an observer flying across the
  horizon $r_h$, would not notice any shocks or peculiarities in the redshift of photons coming to him
  from the shell. Therefore, for the asymptotic behavior (just under the horizon $r_h$) we have
  ${r_{\rm obs^{h-}}\to r_h}$, ${r_{\rm shell^{h-}}\to r_h}$, ${f_{\rm obs^{h-}}\to 0}$, and
  ${f_{\rm shell^{h-}}\to 0}$. On the other hand, according to (\ref{sh-ass7}) and (\ref{sh-dopler8}),
  we have:
\bearr
         {\rm RS_h} = \frac{w_{\rm shell^h}}{w_{\rm obs^h}} = \exp\left[\frac{T_0(r_h - r_c)}{2r_h^2}\right]
\nnn
            \to  \frac{f_{\rm obs^{h-}}}{f_{\rm shell^{h-}}}
           \to \frac{r_{\rm obs^{h-}}-r_h}{r_{\rm shell^{h-}}-r_h},        \label{sh-dopler8_2}
\ear
  whence
\bear
                        r_{\rm shell^{h-}} = r_h + \frac{r_{\rm obs^{h-}}-r_h}{\rm RS_h},    \label{sh-dopler8_3}
\ear
  Therefore, to continue the numerical integration under the horizon ${r_h}$, it is sufficient to take
  a small value of $f_{\rm obs^{h-}} = -f_{\rm obs^{h+}}$ (and the corresponding radius
  $r_{\rm obs^{h-}}$, where the numerical integration was stopped in front of the horizon $r_h$),
  from \eq (\ref{sh-dopler8_3}) we obtain the radius ${r_{\rm shell^{h-}}}$, and with these values, we
  continue the integration, according to its analog in (\ref{sh-ass5_3}):
\bear
                  \int_{r_{\rm obs^{h-}}}^{r_{\rm obs^k}} F_{sum}\, dr
        - \int_{r_{\rm shell^{h-}}}^{r_{\rm shell^k}} F_{sum}\, dr = 0.            \label{sh-ass5_3_T}
\ear
  In addition, it is possible to write an analog of \eq (\ref{sh-ass5}):
\bearr
             T_0 := \int^{r_{\rm obs^{h-}}}_{r_{\rm shell^{h-}}} F_{\rm sum}\, dr
                     = \int^{r_{\rm obs^k}}_{r_{\rm shell^k}} F_{\sum sum}\, dr,
\nnn
         F_{\rm sum}(r\in T) < 0 ,\quad r_{\rm obs^{h-}} < r_{\rm shell^{h-}}. \label{sh-ass5_T}
\ear
  Similarly to the expression (\ref{sh-dopler8_2}), from \eq (\ref{sh-ass5_T}) in the limits
  $r_{\rm obs^k} \to r_c$ and $r_{\rm shell^k} \to r_c$ one can also express the ratio of the radii
  ${r_{\rm obs^{c+}}}$ and ${r_{\rm shell^{c+}}}$ (near the Cauchy horizon $r_c$):
\bearr
             {\rm RS_c} = \frac{w_{\rm shell^c}}{w_{\rm obs^c}} \to \frac{f_{\rm obs^{c+}}}
                    {f_{\rm shell^{c+}}} \to \frac{r_{\rm obs^{c+}}-r_c}{r_{\rm shell^{c+}}-r_c}
\nnn
       \to \exp\left[\frac{-T_0(r_h - r_c)}{2r_c^2}\right] = {\rm RS_h}^{-r_h^2/r_c^2}   \label{sh-ass5_T3}
\ear
  This shows that the frequency shift near the Cauchy horizon $r_c$ for radial photons will is blue-violet,
  i.e., the same as was predicted in the previous section.

\section{Motion under the Cauchy horizon} \label{sh-s_R2}

\begin{figure*}
\centering
\includegraphics[width=0.85\textwidth]{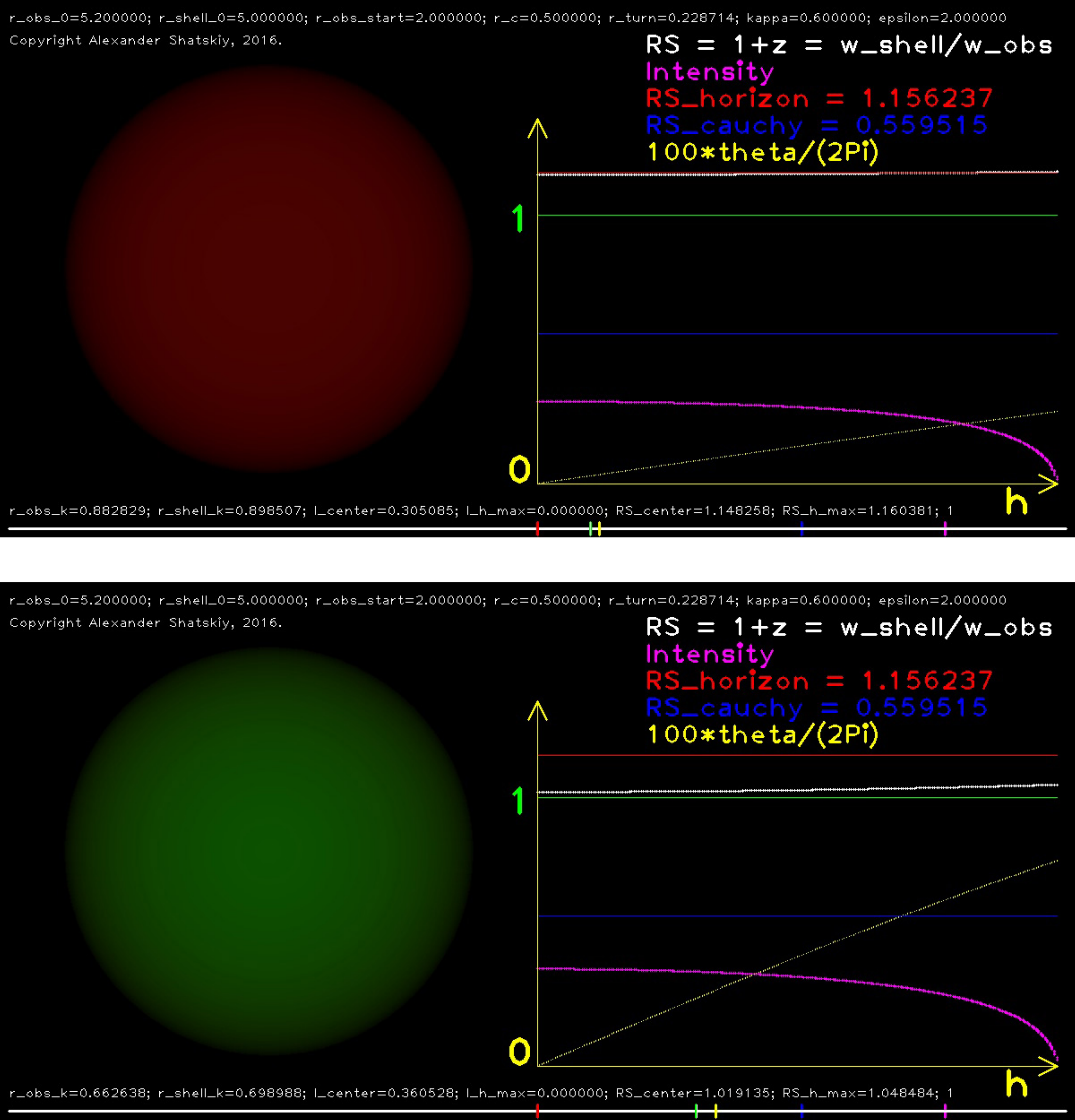}
\caption{\small
        Two video frames for ${r_c=0.5r_h}$, ${\ep = 2}$ at ${r_{\rm obs^k} = 0.41r_h}$ and
        ${r_{\rm obs^k} = 0.38r_h}$, under the Cauchy horizon.} \label{sh-R2}
\end{figure*}
  In the R2-region (under the Cauchy horizon $r_c$) everything is similar to the previous arguments:
\bear
            \int_{r_{\rm obs^{c-}}}^{r_{\rm obs^k}} F_{\rm sum}\, dr
               - \int_{r_{\rm shell^{c-}}}^{r_{\rm shell^k}} F_{\rm sum}\, dr = 0.          \label{sh-ass5_3_3}
\ear
  Similarly to (\ref{sh-ass5_T3}), we obtain
\bear
            \frac{r_c - r_{\rm obs^{c-}}}{r_c - r_{\rm shell^{c-}}} = {\rm RS_c}.       \label{sh-ass5_T3_2}
\ear
  The further integration can be continued up to the ``throat'', the point $r_{\rm turn}$, corresponding
  to the equation $f (r_{\rm turn}) = \ep^2$, see (\ref{sh-U1}):
\bear
       r_{\rm turn} = \frac{r_h+r_c}{2(\ep^2 {-} 1)}\left[\sqrt{1+\frac{4 r_h r_c (\ep^2 {-} 1)}
        {(r_h {+} r_c)^2}} - 1\right].     \label{sh-invers1}
\ear
  After the throat point (in the dynamic Reissner-Nordstr\"om wormhole), matter starts expanding in the
  direction to another universe (in the topology of the total Reissner-Nordstr\"om multiverse,
  similar to what should happen in the Kerr solution). But these effects are beyond thge scope of this paper.

\section{Distribution of energy flux density from the shell}\label{sh-s_intensity}

  The intensity (or more correctly to call it the energy flux density) $I$ is determined by the ratio of the
  photon flux which arriving at the observer (we consider only such photons) to the element of solid angle
  from which these photons are coming per unit natural time of the observer.

  Consider an element of our dust shell in the range of angles from $\theta_{\rm shell}$ to
  ${\theta_{\rm shell} + d\theta_{\rm shell}}$
  (a thin ring with the radius ${2\pi r_{\rm shell}\sin\theta_{\rm shell}}$).
  The photon flux from this element is proportional to ${\sin\theta_{\rm shell}\, d\theta_{\rm shell}}$.
  The solid angle element at the observation point also lies in the range of angles
  from ${\alpha_{\rm obs^k}}$ to ${\alpha_{\rm obs^k} + d\alpha_{\rm obs^k}}$,
  wherein the angle $\alpha_{\rm obs^k}$ corresponds to the angle between the $Z$ axis and
  the direction from which the photon has arrived. Therefore, the solid angle element at the observation point
  is $\sin\alpha_{\rm obs^k}\, d\alpha_{\rm obs^k}$.

  Thus we obtain for the distribution $I$:
\bear
           I(\alpha_{\rm obs^k}) \propto \frac{\sin\theta_{\rm shell}\, d\theta_{\rm shell}/dh}
            {\sin\alpha_{\rm obs^k}\, d\alpha_{\rm obs^k}/dh} \cdot
              \left(\frac{d\tau_{\rm obs^k}}{d\tau_{\rm shell}} \right)^{-2}.         \label{sh-bright1}
\ear
  Here ${d\tau_{\rm obs^k}/d\tau_{\rm shell}}$  is the ratio of the elements of natural time of the observer
  and the shell, which is equal to the ratio of natural frequencies, see RS in (\ref{sh-dopler7}).
  Another RS factor appears in (\ref{sh-bright1}) due to changes in the photon energy while traveling to
  the observer --- it must also be taken into account in the calculation of the energy flux density.
  Therefore, to solve the problem, we need to introduce an invariant definition for the angles between
   the photons' geodesic lines at their points of emission and absorption (separately).

  Since we are interested in the angle between a radial geodesic ($h=0$) and a geodetic which has the
  impact parameter $h >0 $ (at one and the same point), we need the following 4-vector:\footnote
    {The expression for the 4-vector $A_i$ contains the unit antisymmetric (in all four indices) tensor.
      Therefore (strictly speaking), this antisymmetric tensor should also be contracted with the
            antisymmetric kernel tensor\\
       $\Psi^{ml} := [\Psi^m(0)\, \Psi^l(h) - \Psi^l(0)\, \Psi^m(h)]/2$.
            At the contraction, we should take only the antisymmetric part of ${\Psi^m(0)\, \Psi^l(h)}$ since
            the symmetric part of this tensor gives zero.}
\bearr
     A_i : =  E_{ijml}\, U^j\, \Psi^m(0)\, \Psi^l(h),
\nnn
         A^i := E^{ijml}\, U_j\, \Psi_m(0)\, \Psi_l(h) \, . \label{sh-bright2}
\ear
  Her,: ${E_{ijml} = e_{ijml}\sqrt{-g}}$, and ${E^{ijml} = e_{ijml}/\sqrt{-g}}$, a unit, completely
  antisymmetric tensor of rank 4, see [6], \S 83; $g$ is the determinant of our metric tensor;
  ${U^j}$ is the radial velocity 4-vector of the observer or a dust particle of the shell;
  ${\Psi^m(0)}$ is the light 4-vector for a radial photon with ${h=0}$;
   ${\Psi^l(h)}$ is a null 4-vector of a non-radial photon with ${h>0}$.

  It can be seen that in our model the 4-vector $A_i$ has only one nonzero (axial) component
\bear
      A_i = -\delta^\varphi_i \sqrt{-g} [U^t\, \Psi^r(0) - U^r\, \Psi^t(0)] \, \Psi^\theta(h).   \label{sh-bright3_}
\ear
  Similarly, for the contravariant components of the 4-vector ${A^i}$ we have
\bear
          A^i = \frac{-\delta_\varphi^i }{\sqrt{-g}}
                   \left[U_t\, \Psi_r(0) - U_r\, \Psi_t(0) \right] \, \Psi_\theta(h).            \label{sh-bright3}
\ear
  Hence we obtain the scalar $A^2 := A_i A^i$:
\bearr
     A^2 = -\left( U^t\Psi_t(0) + U^r\Psi_r(0) \right)^2 \, \Psi_\theta(h)\Psi^\theta(h)
\nnn
              = -\left( U^j\Psi_j(0) \right)^2 \, \Psi_\theta(h)\Psi^\theta(h).             \label{sh-bright3_2}
\ear
  This shows that it makes sense to introduce one more 4-vector
\bear
                      B_i := \frac{A_i}{U^j\Psi_j(0)}.        \label{sh-bright4}
\ear
  Then for the scalar $B^2: = B_j B^j$ we obtain the following invariant value:
\bear
            B^2 = \frac{\Psi_\theta^2(h)}{r^2} = \frac{h^2\Psi_t^2(h)}{r^2}.            \label{sh-bright5}
\ear
  The value of $h^2/r^2$ by its physical meaning coincides with the square of the sine of the angle
  between the radial and non-radial null geodesics at radius $r$. And the integral of motion $\Psi_t(h)$
  does not change from the point of photon emission to the point of its absorption.
  The invariant expression (\ref{sh-bright5}) is just what we need since it is also true in the comoving,
  freely falling reference frame.

  Then \eq (\ref{sh-bright1}) can be rewritten in an invariant form:
\bear
     I(\alpha_{\rm obs^k}) \propto   {\rm RS}^{-2}\cdot\frac{r_{\rm obs^k}
                \sqrt{r_{\rm obs^k}^2 - h^2}}{r_{\rm shell}\sqrt{r_{\rm shell}^2 - h^2}}.    \label{sh-bright6}
\ear

\section{Boundary effects} \label{sh-s_ring}

  If $h\to r_{\rm shell}$, the energy flux density (\ref{sh-bright6}) on the edges of the visible disk tends
  to infinity. It's not an error! It is due to the fact that near the edge there are many dust rings of our shell
  which emit the photons to the observer. A similar effect can be seen, for example, while observing
  the Earth's atmosphere from space: illuminated by the Sun, the atmosphere at the edge begins to shine
  brighter than at the center and becomes visible. However, if in our model we consider all dust rings
  to be completely opaque for light, then the expression (\ref{sh-bright6}) is multiplied by
  ${\cos\alpha_{\rm shell} = \sqrt{1 - h^2/r_{\rm shell}^2}}$.
  But this is the other extreme. To take into account different options for the transparency of the
  dust rings, we introduce a coefficient ${\kappa \in [0,\, 1]}$ in the expression for $I$:
\bear
           I(\alpha_{\rm obs^k}) \propto {\rm RS}^{-2}\cdot\frac{r_{\rm obs^k}
             \sqrt{r_{\rm obs^k}^2 - h^2}}{r_{\rm shell}\sqrt{r_{\rm shell}^2 - \kappa h^2}}.\label{sh-visual1}
\ear
  Then ${\kappa = 0}$ corresponds to absolute opacity, and ${\kappa = 1}$ to absolute transparency of
  the dust rings.

  The total angle of photon deflection while moving from the shell to the observer is determined
  by the expression (\ref{sh-dopler4}):
\bear           \label{sh-theta_tot}
    \theta_{\rm tot} = \int_{r_{\rm shell}}^{r_{\rm obs^k}} \frac{h \, dr }{r\sqrt{r^2 - h^2 f}}.
\ear
  Thus the maximum possible impact parameter\footnote
          {This maximum possible impact parameter ${h_{\max}}$ is not reached in our case,
       as can be seen from \eq (\ref{sh-visual1}).}
  for $r$ is determined by the relation ${h_{\max} = r/\sqrt{f}}$, and in the T-region there can be any
  values of the impact parameters. For ${h_{\max}}$, the deflection angle of a photon will still be finite,
  despite the (weak) singularity in the denominator (\ref{sh-theta_tot}). This fact is fundamentally different
  from the observation of photons by an observer at rest in the Reissner-Nordstr\"om system.
  In this case, the photons' deflection angles can reach infinity because for sufficiently large values
  of $h$ the photon makes an infinite number of turns around the black hole before it reaches the observer.

\section{Visualization}
\label{sh-s_visuality}

  As the main result of this work, I consider a program (code) due to which it is possible to synthesize
  videos illustrating how the dust shell will look like for a freely falling observer.

  The videos of a Reissner-Nordstr\"om black hole with ${r_c=0.5r_h}$ can be viewed at the addresses:\\
  {https://youtu.be/2PEJp8GOomY} and\\ {https://youtu.be/OiR0X1BwkCo}.

  Videos for a Schwarzschild black hole can be viewed at the following addresses\\
   {https://youtu.be/0gfs-60mlQY} and\\ {https://youtu.be/35VbV6RJgHA}.

  The top and bottom of each video are presented with a string containing all current parameters of this video.

  Visualization for the frequency shift RS on the video frames  requires a detailed explanation.

  The minimum light frequency perceived by human eye corresponds to the red color, and the maximum
  to violet. However, in technical devices (monitors and video projectors), all available colors are obtained
  by mixing the three color channels  RGB (Red, Green and Blue), no violet color among them.
  Meanwhile, the violet color has the highest frequency in this visible light range, to the right of blue,
  therefore, apparently, it cannot be obtained by mixing other colors from the visible range.
  Paradoxically the human eye perceives an artificially synthesized violet color as a mixture in equal
  of red (R) and blue (B) channels (or colors). This fact makes some complexity in video synthesis,
  because the R and B channels correspond to almost opposite ends of the natural color spectrum
  visible by the human eye.

  To make the mapping of colors in the videos maximally relevant to reality, it is reasonable to introduce
  the following assumption in the model: the shell emits only a monochrome line at natural frequency.
  I chose for this monochrome line pure green color (technically, there appears only one green channel, G).

  With regard to the frequency shift to the red region, everything is fairly clear: I assigned the redshift on the
  black hole horizon $r_h$ to corresponds to the red channel R only. Between these two separate
  channels (G and R), there is a smooth frequency change, i.e., the R and G channels provide a smoothly
  varying contribution to the overall color picture.

  For a further frequency shift to the violet region, I have assigned the color on the Cauchy horizon
  as a contribution from the blue channel B only. Between the horizons, the color gradually changes from
  pure red (only the R channel) to pure blue (only B) with a smooth passage through pure green (only the G
  channel, the natural frequency of the shell). Of course, at each point of the dust shell, this smooth color
  change will be of their own, smoothly and continuously connected with adjacent points. A further
  frequency increase occurs beyond the Cauchy horizon (i.e., reduction of the wavelength of light),
  therefore for zero wavelength I assigned an equal contribution from the red (R) and blue (B) channels,
  making the artificially synthesized violet color.

  Of course, this synthetic color scheme is not fully consistent with what the human eye sees in reality,
  but this model maximally transfers all changes in the frequency shift from the monochrome green line
  emitted by dust particles. Moreover, one can observe not only by human eyes but (mostly) by
  technical means (cameras) which are known to work in different frequency bands.

  Regarding visualization of radiation intensity, it is clear that for changing the intensity it is sufficient
  to synchronously change the contributions to the share in all channels (RGB), ranging from zero to
  255 (maximally technically possible value).

  On the right-hand side of the frame of all videos there are plots which synchronously change in time
  with the main picture: for the redshift values (the white curve), intensity (the violet curve) and
  the total deflection angle of photons (yellow curve), depending on the impact parameter $h$ of the photons.

  In addition, the three horizontal solid lines are shown on the frames: the red one for the RS value on the
  event horizon $r_h$, the green one indicating the unit, and the blue one for the RS value on the
  Cauchy horizon.

  At the bottom, the scale of the observer radius is displayed (marked with a yellow label),
  that for the point of the dust shell nearest to the observer (a green label) and its maximally remote
  from the observer from which photons are coming to the observer (a white label), as well as the
  location of the event horizon (red label), the Cauchy horizon (blue label) and the throat (violet label).

\section{Conclusion}\label{sh-s_conclusion}

  All previous analytical arguments are confirmed by the numerical results displayed in the videos.
  These results show that the observer's trip through both horizons and through the throat point
  ${r_{\rm turn}}$ proceeds smoothly and continuously, in particular, in the sense of observation
  of different elements of the dust shell.

\subsection*{Acknowledgments}\label{sh-s_acknowledgments}

  I am sincerely grateful to Alexei Toporensky and Oleg Zaslavskii for all ideas, comments and
  discussions at the stage of preparation of this paper.

\end{document}